\documentclass{article}
\usepackage{waspaa23,amsmath,url,times}
\usepackage{graphicx}
\usepackage{color}

\usepackage{ifthen}
\usepackage[colorlinks=false,pdfborder={0 0 0}]{hyperref} 
\usepackage{units}
\usepackage{tikz}
\usepackage{amssymb}
\usepackage[subtle]{savetrees}

\usepackage{cite}               
\usepackage{balance} 

\usepackage{caption}
\usepackage{subcaption}

\usepackage{epstopdf}

\usepackage{blindtext}

\title{Relative Transfer Function Vector Estimation for Acoustic Sensor\\ Networks Exploiting Covariance Matrix Structure}

\name{Author 1,$^{1}$
      Author 2,$^{2}$
      Author 3,$^{3}$
      Author 4$^{2}$}
\address{$^1$ Author names and affiliations omitted for double-blind review.\\
         $^2$ Please do not make changes to this section.\\
         $^3$ Author information may be added after paper acceptance.\\
}

\name{Wiebke Middelberg, Henri Gode, Simon Doclo\thanks{This work was funded by the Deutsche Forschungsgemeinschaft (DFG, German Research Foundation) - Project ID 352015383 (SFB 1330 B2) and Project ID 390895286 (EXC 2177/1).
}}
\address{Department of Medical Physics and Acoustics and Cluster of Excellence Hearing4all,\\ University of Oldenburg, Germany, 
\{\href{mailto:wiebke.middelberg@uni-oldenburg.de}{wiebke.middelberg}, henri.gode, simon.doclo\}@uni-oldenbrug.de
}

\makeatletter
\newsavebox\myboxA
\newsavebox\myboxB
\newlength\mylenA
\newcommand*\xoverline[2][0.75]{%
    \sbox{\myboxA}{$\m@th#2$}%
    \setbox\myboxB\null
    \ht\myboxB=\ht\myboxA%
    \dp\myboxB=\dp\myboxA%
    \wd\myboxB=#1\wd\myboxA
    \sbox\myboxB{$\m@th\overline{\copy\myboxB}$}
    \setlength\mylenA{\the\wd\myboxA}
    \addtolength\mylenA{-\the\wd\myboxB}%
    \ifdim\wd\myboxB<\wd\myboxA%
       \rlap{\hskip 0.5\mylenA\usebox\myboxB}{\usebox\myboxA}%
    \else
        \hskip -0.5\mylenA\rlap{\usebox\myboxA}{\hskip 0.5\mylenA\usebox\myboxB}%
    \fi}

\newcommand{\y}{\mathbf{y}}
\newcommand{\x}{\mathbf{x}}
\newcommand{\n}{\mathbf{v}}
\newcommand{\R}{\mathbf{R}}
\newcommand{\Ry}{\R_y}
\newcommand{\Ryest}{\hat{\R}_y}
\newcommand{\Ryestw}{\Ryest^{\rm w}}
\newcommand{\Rx}{\R_x}
\newcommand{\Rxest}{\hat{\R}_x}
\newcommand{\Rn}{\R_v}
\newcommand{\Rnest}{\hat{\R}_v}
\newcommand{\Rnmodest}{\hat{\underline{\R}}_v}
\newcommand{\Rnn}[1]{\R_{v,#1}}
\newcommand{\Rnnest}[1]{\hat{\R}_{v,#1}}
\newcommand{\Ryn}[1]{\R_{y,#1}}
\newcommand{\Rxn}[1]{\R_{x,#1}}
\newcommand{\Smat}{\mathbf{S}}
\newcommand{\evec}{\mathbf{e}}
\newcommand{\h}{\mathbf{h}}
\newcommand{\hest}{\hat{\mathbf{h}}}
\newcommand{\hestnn}{\hest'}
\newcommand{\inv}{^{-1}}
\newcommand{\w}{\mathbf{w}}
\newcommand{\phix}{\phi_x}

\newcommand{\vmax}{\mathbf{v}_{\mathrm{max}}}
\newcommand{\sigmamax}{\sigma_{\mathrm{max}}}
\newcommand{\refsub}{\mathrm{ref}}
\newcommand{\zeros}[2]{\mathbf{0}_{#1\times#2}}
\newcommand{\ones}[2]{\mathbf{1}_{#1\times#2}}

\DeclareMathSymbol{\shortminus}{\mathbin}{AMSa}{"39}

\begin{document}

\ninept
\maketitle

\begin{sloppy}

\begin{abstract}

In many multi-microphone algorithms for noise reduction, an estimate of the relative transfer function (RTF) vector of the target speaker is required.
The state-of-the-art covariance whitening (CW) method estimates the RTF vector as the principal eigenvector of the whitened noisy covariance matrix, where whitening is performed using an estimate of the noise covariance matrix. In this paper, we consider an acoustic sensor network consisting of multiple microphone nodes. Assuming uncorrelated noise between the nodes but not within the nodes, we propose two RTF vector estimation methods that leverage the block-diagonal structure of the noise covariance matrix. The first method modifies the CW method by considering only the diagonal blocks of the estimated noise covariance matrix. In contrast, the second method only considers the off-diagonal blocks of the noisy covariance matrix, but cannot be solved using a simple eigenvalue decomposition. When applying the estimated RTF vector in a minimum variance distortionless response beamformer, simulation results for real-world recordings in a reverberant environment with multiple noise sources show that the modified CW method performs slightly better than the CW method in terms of SNR improvement, while the off-diagonal selection method outperforms a biased RTF vector estimate obtained as the principal eigenvector of the noisy covariance matrix.

\end{abstract}

\begin{keywords}
Acoustic sensor networks, relative transfer function vector, beamforming, covariance whitening
\end{keywords}

\section{Introduction}\label{sec:intro}

Acoustic sensor networks (ASNs) with multiple spatially distributed microphone nodes are of rising interest for speech communication applications due to their ability to capture spatially diverse information \cite{Bertrand2011}. This allows ASNs to be deployed, e.g., for speech enhancement \cite{MarkovichGolan2015ASN,Tavakoli2016ASN,Koutrouvelis2018WASN,Zhang2019RTF,GoesslingWASPAA2019,Corey2021}, and sound source localization \cite{Cobos2017survey}, 
in applications such as smart speakers or hearing aids connected with external microphones. In these applications the desired speech signal is often corrupted by background noise.
To achieve noise reduction, multi-microphone algorithms like the minimum variance distortionless response (MVDR) beamformer can be used \cite{MarkovichGolan2015ASN,GoesslingWASPAA2019,Doclo2015,Gannot2017,VanVeen1988a}, requiring an estimate of the noise covariance matrix and the relative transfer function (RTF) vector of the target speaker. 

In this paper, we consider an ASN where the noise component between all nodes is assumed to be uncorrelated, which is for example the case in a diffuse noise field when the distance between the nodes is large or when different nodes capture different noise sources. This results in a block-diagonal structure of the noise covariance matrix. Exploiting this covariance matrix structure, we propose two methods to estimate the RTF vector of the target speaker in an ASN with at least three nodes. The first method involves a modification of the state-of-the-art covariance whitening (CW) method \cite{Markovich2009,Serizel2014,Markovich2015}.
Instead of using the entire estimated noise covariance matrix for whitening, the proposed CW-D method considers only the diagonal blocks, which allows for efficient inversion and square-root decomposition. The CW and CW-D methods both estimate the RTF vector as the best rank-1 approximation of the whitened noisy covariance matrix, which can be achieved via an eigenvalue decomposition (EVD).

The second method only requires the noisy covariance matrix and no estimate of the noise covariance matrix. Assuming uncorrelated noise between nodes, all information required for RTF vector estimation is contained in the off-diagonal blocks of the noisy covariance matrix. In the off-diagonal selection (ODS) method, an optimization problem is formulated to estimate the entire RTF vector using only the internode correlations of the noisy covariance matrix. Contrary to the first method, the solution of this optimization problem cannot be computed via an EVD, and we propose to use an iterative optimization procedure.

In the experimental evaluation with reverberant real-world recordings and multiple noise sources, the performance is evaluated in terms of RTF vector estimation accuracy and signal-to-noise ratio (SNR) improvement when applying the RTF vector estimates in an MVDR beamformer. The results show that the proposed CW-D method performs slightly better than the CW method. In addition, the proposed ODS method outperforms a biased estimator using the EVD of the entire noisy covariance matrix, especially at low input SNRs where the influence of noise on the diagonal blocks is most severe.

\section{Signal Model and Notation}\label{sec:Signal}

We consider an ASN with $N$ spatially distributed nodes, where node $n\in \{1,\dots,N\}$ contains $M_n$ microphones, i.e., in total $M = \sum_{n=1}^N M_n$ microphones. The considered acoustic scene consists of a single target speaker and undesired ambient noise. The noisy $m$-th microphone signal of the $n$-th node can be written in the short-time Fourier transform (STFT) domain as
\begin{equation}\label{eq:SigSum}
    Y_{n,m}(k,l) = X_{n,m}(k,l) + V_{n,m}(k,l)\, ,
\end{equation}
where $k$ is the frequency bin index and $l$ is the frame index, which - for the sake of brevity - are omitted in the remainder of this paper wherever possible. The speech and noise signal components are denoted by $X_{n,m}$ and $V_{n,m}$, respectively.
The $M_n$-dimensional signal vector for the $n$-th node is defined as
\begin{equation}\label{eq:SigVecNode}
    \y_n = [Y_{n,1},\; Y_{n,2},\;\dots\; ,\; Y_{n,M_n}]^T\, ,    
\end{equation}
where $\{\cdot\}^T$ denotes the transpose operator.
By stacking all node-wise signal vectors, the $M$-dimensional signal vector $\y$, containing all microphone signals in the ASN, is defined as
\begin{equation}\label{eq:SigVec}
    \y = [\y_1^T,\;\y_2^T,\;\dots\; ,\;\y_N^T]^T\, .
\end{equation}
The speech vectors $\x_n$ and $\x$ and the noise vectors $\n_n$ and $\n$ are defined similarly to \eqref{eq:SigVecNode} and \eqref{eq:SigVec}, respectively.
For the speech component, we assume a multiplicative transfer function model \cite{Avargel2007multiplicative}, allowing the speech vector to be written as 
\begin{equation}\label{eq:xvec}
    \x = \h X_\refsub\, ,
\end{equation}
where $\h\in\mathbb{C}^M$ is the target RTF vector, which relates the speech component in the reference microphone $X_\refsub$ to the speech component in all other microphones. Hence, the entry of $\h$ corresponding to the reference microphone is equal to 1. 
It should be noted that the reference microphone is chosen for the entire ASN and not per node.
The RTF vector for node $n$ is defined as $\h_n\in\mathbb{C}^{M_n}$, such that $\h = [\h_1^T,\h_2^T,\dots,\h_N^T]^T$, where all $\h_n$ are normalized to the same reference. 

Assuming that the speech and noise signals are mutually uncorrelated, the noisy covariance matrix $\Ry$ can be written in terms of the speech covariance matrix $\Rx$ and the noise covariance matrix $\Rn$ as
\begin{equation}\label{eq:Ry}
    \Ry = \mathcal{E}\{\y\y^H\} = \Rx +\Rn\, ,
\end{equation}
where $\mathcal{E}\{\cdot\}$ denotes the expectation operator and $\{\cdot\}^H$ denotes the Hermitian transpose operator. The node-wise covariance matrices for the $n$-th node, $\Ryn{n}$, $\Rxn{n}$ and $\Rnn{n}$, are defined similarly to \eqref{eq:Ry}, using the node-wise vectors $\y_n$, $\x_n$ and $\n_n$, respectively.
Using \eqref{eq:xvec}, the speech covariance matrix $\Rx$ can be written as a rank-1 matrix spanned by the RTF vector $\h$, i.e.,
\begin{equation}\label{eq:Rx}
    \Rx = \phix \h\h^H\, ,
\end{equation}
where $\phix = \mathcal{E}\{|X_\refsub|^2\}$ denotes the speech power spectral density (PSD) in the reference microphone.

In this paper, we make the central assumption that the noise component is uncorrelated between different nodes. 
This assumption is realistic, e.g., for a diffuse noise field when the distance between the nodes is large enough \cite{GoesslingWASPAA2019,Corey2021,Middelberg2022IWAENC} or when nodes capture different noise sources.
For the noise correlation between different microphones within each node, no assumption is made, implying that within each node the noise component may be partially correlated. The node-wise noise covariance matrices $\Rnn{n}$ are assumed to be full-rank.
Figure \ref{fig:CovMat} schematically depicts the structure of the entire noise covariance matrix $\Rn$, where yellow indicates high correlation (within each node), and white indicates low correlation (between the nodes).
To visualize the influence of such a  block-diagonal noise covariance matrix on the noisy covariance matrix, Figure \ref{fig:CovMat} also depicts the rank-1 speech covariance matrix $\Rx$ and the resulting noisy covariance matrix $\Ry$. It can clearly be seen that $\Ry$ only contains information about the target RTF vector in its off-diagonal blocks (orange and green), as they are unaffected by the noise. 
Note that for $N=2$ nodes, the off-diagonal block of $\Ry$ only contains information about scaled versions of $\h_1$ and $\h_2$, which cannot be unified into the vector $\h=[\h_1^T,\h_2^T ]^T$, as the speech PSD $\phix$ (see \eqref{eq:Rx}) evokes a scaling ambiguity for the RTF vector part that does not contain the reference microphone. This scaling ambiguity, however, can be lifted when $N\geq 3$, since direct information about the relative scaling of the different parts is contained in adjacent off-diagonal blocks.




To achieve noise reduction, we consider the MVDR beamformer, which requires an estimate of the noise covariance matrix $\Rnest$ and an estimate of the RTF vector $\hest$. The filter vector $\w$ of the MVDR beamformer is given by \cite{Gannot2017,Doclo2015}
\begin{equation}\label{eq:MVDR}
    \w = \frac{\Rnest\inv\hest}{\hest^H\Rnest\inv\hest}\, ,
\end{equation}
yielding the output signal $Z = \w^H\y$ when applied to the noisy input signals. The filtered speech and noise components are defined as $Z_x = \w^H\x$ and $Z_v = \w^H\n$, respectively.

\begin{figure}[t]
  \centering
  \includegraphics[width=0.95\linewidth]{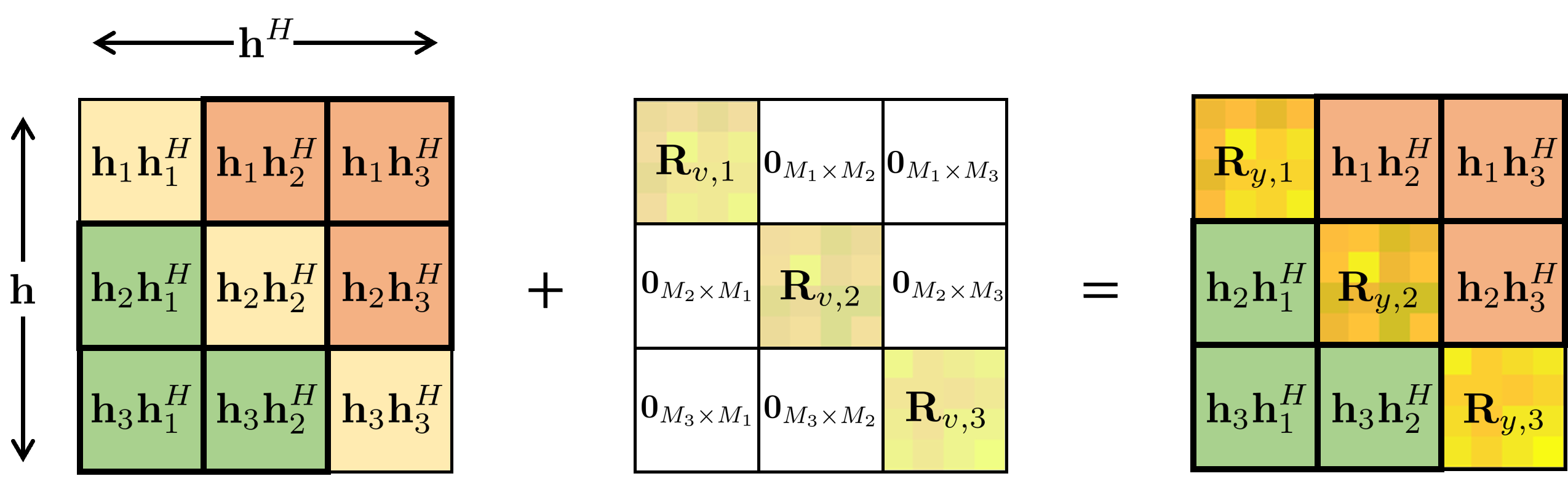}
  \vspace{-0.1cm}
  \caption{Visualization of the influence of noise on the noisy covariance matrix for $N$ = 3 nodes (with $\phix$ = 1). Left: Speech covariance matrix spanned by the target RTF vector. Middle: Noise covariance matrix, uncorrelated between nodes. Right: Noisy covariance matrix, containing only information about the target RTF vector on its off-diagonal blocks.}
  \vspace{-0.3cm}
  \label{fig:CovMat}
\end{figure}

\section{RTF Vector Estimation Methods}\label{sec:RTF}

In this section, we present different RTF vector estimation methods, where we first discuss the general idea of the rank-1 approximation (Section \ref{subsec:R1}), which is the basis for the biased estimator and the state-of-the-art CW method (Section \ref{subsec:CW}). In Section \ref{subsec:CWB}, we propose a modification of the CW method by exploiting the assumed block-diagonal structure of the noise covariance matrix. In Section \ref{subsec:SORY}, a novel cost function is proposed, where only the off-diagonal blocks of the noisy covariance matrix are selected to compute the best fitting RTF vector.

\subsection{Rank-1 Approximation and Biased Estimator}\label{subsec:R1}

To motivate the RTF vector estimation methods in the following sections, we first consider the case where an estimate of the speech covariance matrix $\Rxest$ is available. In practice, the rank-1 model in \eqref{eq:Rx} does not perfectly hold, e.g., due to an insufficient frame length. Hence, the RTF vector $\hest$ is often estimated as the best rank-1 approximation of $\Rxest$ \cite{Serizel2014}, i.e., the vector solving the optimization problem
\begin{equation}\label{eq:CostOpt}
    \min_{\hestnn} \underbrace{\| \Rxest - \hest'\hest'^H \|_F^2}_{J(\hest')}\, ,
\end{equation}
where $\hestnn$ is a scaled (non-normalized) version of $\hest$ and $\|\cdot\|_F$ denotes the Frobenius norm.
Considering the gradient of the cost function $J(\hest')$ in \eqref{eq:CostOpt}, i.e.,
\begin{equation}\label{eq:GradEVD}
    \nabla J(\hestnn) = -2\, \left(\Rxest-\hestnn\hestnn^H\right)\hestnn\, ,
\end{equation}
and setting it to zero, the solutions can be found by solving the eigenvalue problem
\begin{equation}\label{eq:EigProb}
    \Rxest\hestnn = \hestnn(\hestnn^H\hestnn)\, .
\end{equation}
Hence, the best rank-1 approximation of $\Rxest$ is a scaled version of the principal eigenvector $\vmax$, corresponding to the maximum eigenvalue $\sigmamax$. The RTF vector estimate can then be obtained as
\begin{equation}\label{eq:hestvmax}
    \hest = \frac{\vmax}{\mathbf{e}_\refsub^T\vmax}\, ,
\end{equation}
where $\evec_\refsub$ is a selection vector containing all zeros except for the entry corresponding to the reference microphone, which equals 1.

If only the noisy covariance matrix $\Ryest$ is available, a biased estimate may be obtained as the best rank-1 approximation of $\Ryest$, i.e.,
\begin{equation}\label{eq:OptProbEVD}
    \min_{\hestnn} \| \Ryest - \hestnn\hestnn^H \|_F^2\, ,
\end{equation}
where the bias obviously is larger for lower SNR. The biased RTF vector estimate $\hest^{\rm B}$ is obtained as in \eqref{eq:hestvmax}, where $\vmax$ is the principal eigenvector of $\Ryest$.

\subsection{Covariance Whitening (CW)}\label{subsec:CW}

{To compensate for the influence of the ambient noise on the RTF vector estimate, a frequently used approach is to perform whitening of the noisy covariance matrix using an estimate of the noise covariance matrix, i.e., $\Ryestw = \Rnest^{-1/2}\Ryest\Rnest^{-H/2}$, where $\Rnest = \Rnest^{1/2}\Rnest^{H/2}$ corresponds to a square-root decomposition, e.g., the Cholesky decomposition \cite{Markovich2009,Serizel2014,Markovich2015}.\parfillskip=0pt\par}
Similarly to \eqref{eq:CostOpt}, the optimization problem in the whitened domain is given by
\begin{equation}\label{eq:CWcost}
    \min_{\hestnn} \| \Ryestw - \Rnest^{-1/2}\hestnn\hestnn^H\Rnest^{-H/2} \|_F^2\, ,
\end{equation}
i.e., the principal eigenvector $\vmax$ of the whitened noisy covariance matrix $\Ryestw$ corresponds to a scaled version of the whitened RTF vector estimate. By de-whitening and normalizing $\vmax$, the CW RTF vector estimate is obtained as
\begin{equation}\label{eq:hCW}
    \hest^{\rm CW} = \frac{\Rnest^{1/2}\vmax}{\evec_\refsub^T\Rnest^{1/2}\vmax}\, .
\end{equation}

\subsection{Covariance Whitening Using Diagonal Blocks (CW-D)}\label{subsec:CWB}

To leverage the assumed block-diagonal structure of the noise covariance matrix (see Figure \ref{fig:CovMat}), we propose to only consider the diagonal blocks of $\Rnest$ in the CW method. The modified noise covariance matrix $\Rnmodest$ is constructed as a block-diagonal matrix containing the node-wise noise covariance matrices $\Rnnest{n}$ on its diagonal blocks (and all zeros in the off-diagonal blocks).
For a block-diagonal matrix, the inverse and the square-root decomposition can be performed efficiently on the separate diagonal blocks, such that the matrix required for the whitening operation in \eqref{eq:CWcost} is given by
\begin{equation}\label{eq:RnInvBlock}
    \Rnmodest^{-1/2} = \left[\begin{array}{cccc}
        \Rnnest{1}^{-1/2} &  \zeros{M_1}{M_2} & \dots & \zeros{M_1}{M_N}\\
        \zeros{M_2}{M_1} & \Rnn{2}^{-1/2} & \dots & \vdots\\
        \vdots & \vdots & \ddots & \vdots\\
        \zeros{M_N}{M_1} & \dots & \dots & \Rnnest{N}^{-1/2}\\
    \end{array}\right]\, .
\end{equation}
The whitened RTF vector estimate is obtained as the principal eigenvector of $\Rnmodest^{-1/2}\Ryest\Rnmodest^{-H/2}$, where de-whitening and normalization is performed similarly to \eqref{eq:hCW} using $\Rnmodest^{1/2}$ to obtain the RTF vector estimate $\hest^{\rm CW\shortminus D}$.


\subsection{Off-Diagonal Selection (ODS)}\label{subsec:SORY}

In the optimization problem for the biased estimator in \eqref{eq:OptProbEVD}, it can directly be seen that biased information is used, as the diagonal blocks of $\Ryest$ contains both speech and noise information. To avoid this problem without compensating for the noise directly (as in the CW and CW-D methods), we propose the following optimization problem
\begin{equation}\label{eq:OptProbSORY}
    \min_{\hestnn} \underbrace{\| \Smat\odot \left(\Ryest -\hestnn\hestnn^H \right)\|_F^2}_{J(\hestnn)}\, ,
    \vspace{-0.2cm}
\end{equation}
where we only select the off-diagonal blocks of $\Ryest$ (see Figure \ref{fig:CovMat}) and its respective rank-1 approximation by means of the selection matrix $\Smat$ and $\odot$ denotes the Hadamard product, i.e., the element-wise multiplication of matrices. The selection matrix is defined as

\begin{equation}\label{eq:Smat}
    \Smat = \left[\begin{array}{cccc}
        \zeros{M_1}{M_1} &  \ones{M_1}{M_2} & \dots & \ones{M_1}{M_N}\\
        \ones{M_2}{M_1} & \zeros{M_2}{M_2} & \dots & \vdots\\
        \vdots & \vdots & \ddots & \vdots\\
        \ones{M_N}{M_1} & \dots & \dots & \zeros{M_N}{M_N}\\
    \end{array}\right]\, ,
\end{equation}
containing all ones except for the diagonal blocks, which contain zeros.

\noindent
Similarly to \eqref{eq:GradEVD}, the gradient of the cost function $J(\hest')$ in \eqref{eq:OptProbSORY} is given by
\begin{equation}\label{eq:GradSORY}
    \nabla J(\hestnn) = -2\, \left(\Smat\odot\left(\Ryest-\hestnn\hestnn^H\right)\right)\hestnn \, .
\end{equation}
Setting the gradient equal to zero yields
\begin{equation}\label{eq:SORYHada}
    (\Smat\odot\Ryest)\hestnn = (\Smat\odot(\hestnn\hestnn^H))\hestnn\, .
\end{equation}
In contrast to \eqref{eq:EigProb}, this does not correspond to an eigenvalue problem, since the Hadamard product does not allow for further simplification.
To the best of our knowledge, there is no closed-form solution or well-defined operation like the (generalized) EVD to solve \eqref{eq:SORYHada}. 
Nevertheless, iterative optimization procedures like gradient-descent or the quasi-Newton method can be used \cite{Fletcher1987Optimization}.
After solving the unconstrained optimization problem in \eqref{eq:OptProbSORY}, the RTF vector estimate $\hest^{\rm ODS}$ is obtained by normalizing the solution to the reference entry.

In general, it should be noted that the optimization problems for the ODS method in \eqref{eq:OptProbSORY} and the biased estimator in \eqref{eq:OptProbEVD} only require the noisy covariance matrix $\Ryest$, whereas the optimization problem for the CW and CW-D methods in \eqref{eq:CWcost} also require an estimate of the noise covariance matrix $\Rnest$.

\section{Evaluation}\label{sec:eval}
In this section, we evaluate the performance of the presented RTF vector estimation methods using real-world recordings. The considered performance measures are the Hermitian angle between the ground truth RTF vector and the estimated RTF vector, and the intelligibility-weighted SNR improvement of the MVDR beamformer using the respective RTF vector estimates.

\subsection{Setup and Implementation}\label{subsec:Setup}

Figure \ref{fig:SceneEval} depicts the considered acoustic scene for the evaluation, consisting of a target speaker in an acoustically treated laboratory with dimensions 7$\times$6$\times$2.7 $\rm m^3$ and a reverberation time $T_{60}\approx$ 500 ms. As speech material, four different talkers (two male and two female) from the EBU SQAM CD \cite{ebuSQAM2} and the VCTK corpus \cite{Veaux2017VCTKmics} were used. 
All utterances had a duration of 20 s. The ambient noise was generated by four loudspeakers in the corners of the room. 
Different versions of multi-talker babble noise were played back at approximately the same level by the four loudspeakers.

The acoustic sensor network for the evaluation consisted of four nodes with uniform linear arrays, placed at about 0.5 m distance from the noise loudspeakers. 
Nodes 1-3 contained four microphones each, while node 4 contained three microphones, giving a total of $M$=15 microphones. For all arrays, two different microphone spacings of 1 cm and 3 cm were considered. The first microphone of node 1 was chosen as the reference microphone.
The speech and noise components were recorded separately at a sampling rate of 16 kHz and mixed subsequently at an SNR of $\rm SNR_{in}=\{-5,\, 0,\, 5\}$ dB in the reference microphone.

\begin{figure}
  \centering
  \includegraphics[width=0.89\linewidth]{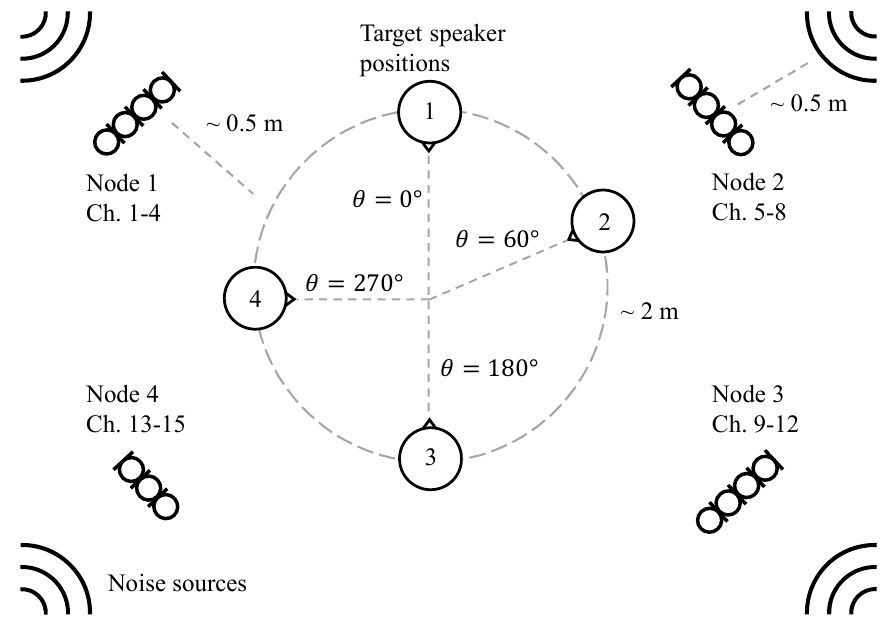}
  \vspace{-0.2cm}
  \caption{Schematic setup of acoustic scene with four different speaker target positions, four microphone nodes and four distinct noise sources.}
  \label{fig:SceneEval}
  \vspace{-0.2cm}
\end{figure}


For the implementation of the algorithms, an STFT framework with a frame length of 512 samples (corresponding to 32 ms), a frame overlap of 50\%, and a square-root-Hann window for analysis and synthesis was used. The covariance matrices were estimated in batch, where for each frequency bin $\Ryest$ was estimated during speech activity and $\Rnest$ was estimated during speech pauses. For each frequency bin, speech-plus-noise and noise-only frames were determined by means of a speech presence probability (SPP) estimator \cite{Gerkmann2012}, which was computed on one microphone per node and subsequently averaged.


The MVDR beamformer was computed according to \eqref{eq:MVDR}, where the estimated noise covariance matrix was used in conjunction with one of the four presented RTF vector estimates:
\begin{itemize}
    \item CW: EVD of $\Ryest$ whitened with $\Rnest$.
    \item CW-D: CW using block-diagonal noise covariance matrix $\Rnmodest$.
    \item Biased estimator: EVD of $\Ryest$.
    \item ODS: Iterative optimization method using only off-diagonal blocks of $\Ryest$. Optimized using \textsc{Matlab}'s \texttt{fminunc} function \cite{MatlabOptimization} supplied with the gradient and initialized on a random vector.
\end{itemize}

As a measure of RTF vector estimation accuracy, we use the Hermitian angle \cite{Varzandeh2017} 
between the ground truth RTF vector $\h$ and the estimated RTF vector $\hest$, i.e., $\theta = \arccos\left( \frac{|\h^H\hest|}{\|\h \|_2 \|\hest \|_2 } \right)$, averaged over all frequency bins. For each target position, the ground truth RTF vector was computed via the EVD of the oracle speech covariance matrix, obtained using the measured room impulse response convolved with white Gaussian noise.
The second performance measure is the intelligibility-weighted SNR improvement \cite{Greenberg1993iSNR} $\Delta \rm SNR =  SNR_{out} - SNR_{in,max}$, where the output SNR $\rm SNR_{out}$ is computed using the filtered speech and noise signal components and $\rm SNR_{in,max}$ is the highest input SNR among all microphones.


\subsection{Results and Discussion}\label{subsec:results}

\begin{figure}[t]
  \centerline{\includegraphics[width=0.92\columnwidth,trim={1cm 0cm 5.5cm 18.8cm},clip]{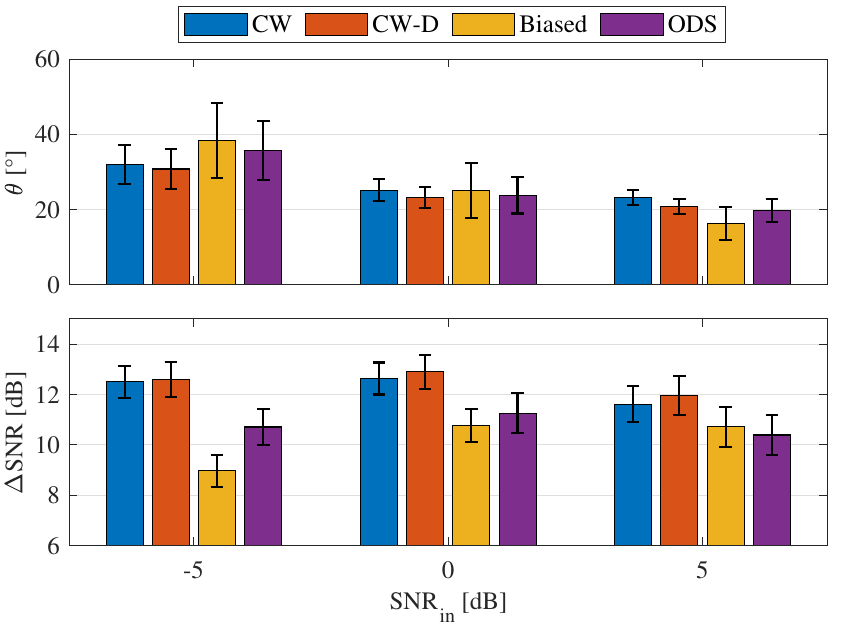}}
  \vspace{-0.4cm}
  \caption{Hermitian angle between ground truth RTF vector and different RTF vector estimates (upper panel) and SNR improvement (lower panel) for different input SNRs.}
  \label{fig:Results}
  \vspace{-0.2cm}
\end{figure}

For different input SNRs, Figure \ref{fig:Results} depicts the Hermitian angle and the SNR improvement for the considered RTF vector estimation methods, where the bars represent the mean over 32 conditions (four target speaker positions, four speakers, two microphone spacings) and the error bars depict the standard deviation.
First, it can be observed that in general the Hermitian angle in the upper panel of Figure \ref{fig:Results} decreases with increasing input SNR, implying more accurate estimation at higher SNRs. Although at low SNRs the CW and CW-D methods achieve a lower Hermitian angle than the biased estimator and ODS method, these differences become negligible at higher input SNRs. 

Second, it can be observed that in terms of SNR improvement the differences between the methods are more noticeable than in terms of Hermitian angle. At all input SNRs, the CW and CW-D methods consistently achieve around 12 dB of SNR improvement and outperform the biased estimator and the ODS  method. This indicates a clear benefit of compensating for the noise using the estimated noise covariance matrix $\Rnest$ instead of using biased or selected information from the covariance matrix $\Ryest$.
The performance of the CW and CW-D methods is similar, although the CW-D method attains a slightly higher SNR improvement, particularly at higher input SNRs. These results indicate a good validity of the block-diagonal model for the noise covariance matrix for the considered scenario. Hence, inverting only the diagonal blocks, cf. \eqref{eq:RnInvBlock}, seems to be sufficient or even beneficial, as it may reduce estimation errors of $\Rnest$.

Comparing the SNR improvement of the biased estimator with the ODS method, it can be observed that at an input SNR of -5 dB, the ODS method significantly outperforms the biased estimator (by about 1.5 dB). At higher input SNRs, the advantage of using only the off-diagonal blocks vanishes, and it seems more beneficial to use all information as the influence of noise diminishes. This indicates a higher robustness of the EVD of the full covariance matrix towards deviations from the rank-1 speech model compared to selecting only the off-diagonal blocks. At low SNRs, however, it seems more advantageous to exclude biased information and use only the off-diagonal blocks, which are affected less by noise, leading to a better performance of the ODS method compared to the biased estimator.



\section{Conclusion}

In this paper, we presented and compared different RTF vector estimation methods leveraging the assumed block-diagonal structure of the noise covariance matrix in an acoustic sensor network with multiple nodes. In an evaluation with real-world recordings, the modified CW method, which only considers the diagonal blocks of the noise covariance matrix, showed equal or even slightly better results than the original CW method at a lower complexity.  Furthermore, we proposed a novel optimization problem for RTF vector estimation by selecting only the off-diagonal blocks of the noisy covariance matrix which are assumed not to be affected by noise. The evaluation results showed that the ODS method clearly outperforms a biased estimator in terms of SNR improvement, especially at low SNRs, showing that the selection of only unbiased information is beneficial if the influence of noise is large.

\bibliographystyle{IEEEtran}
\interlinepenalty=10000 
\bibliography{main}

\end{sloppy}
\end{document}